# A Novel Taxonomy and Classification Scheme for Code Smell Interactions

**Ruchin Gupta[1], Sandeep Kumar Singh[2]**
[1,2]Jaypee Institute of Information Technology, Noida,
[1]skg11in@yahoo.co.in
[2]sandeepk.singh@jiit.ac.in

**Abstract**

Code smells are indicators of potential design flaws in source code and do not appear alone but in combination with other smells, creating complex interactions. While existing literature classifies these smell interactions into collocated, coupled, and inter-smell relations, however, to the best of our knowledge, no research has used the existing knowledge of code smells and (or) their relationships with other code smells in the detection of code smells. This gap highlights the need for deeper investigation into how code smells interact with each other and assist in their detection. This would improve the overall comprehension of code smells and how they interact more effectively. This study presents a novel taxonomy and a proposed classification scheme for the possible code smell interactions considering a specific programming language as a domain. This paper has dealt with one scenario called Inter smell detection within the domain. The experiments have been carried out using several popular machine learning (ML) models. Results primarily show the presence of code smell interactions namely Inter-smell Detection within domain. These results are compatible with the available facts in the literature suggesting a promising direction for future research in code smell detection.

## 1. Introduction & background

Kent Beck introduced the concept of "*code smell*" in 1999[1], describing it as certain patterns in code that suggest or strongly indicate a need for refactoring. Code smells are signs of potential issues that could undermine the software's quality attributes [1][2]. When numerous code smells accumulate, they can greatly impair the software's maintainability. Martin Fowler initially provided a list of 22 distinct types of code smells, which were later categorized into groups such as *Bloaters, Object Orientation Abusers, Change Preventers, Dispensables, Couplers, and Encapsulators* [3]. The latest study by Rasool et al. [4] proposed a scientific classification of these 22 code smells into five distinct categories, aimed at improving understanding by grouping them based on shared characteristics like improper behavior and poor organization of code components, etc.

The classifications have sparked discussions about the potential relationships between code smells and their impact on various aspects of the code being analyzed. According to a literature review, code smells do not occur in the code alone [5, 6] and is accompanied by other code smells [7–11]. Thus, code smells interact, and the existing literature has explored code smell interactions in terms of collocation, coupling [9, 12–14], and inter-smell relationships [15].

Also, the literature has placed significant emphasis on detecting code smells through a variety of techniques. Existing literature reveals extensive efforts to identify various code smells [16–19]. Techniques for detecting code smells are generally categorized into five main strategies: Metric-based, Rule/Heuristic-based, History-based, Optimization-based, and Machine Learning-based. Metric-based Strategy [20–22] identifies code smells by analyzing metrics and their threshold values. Rule or Heuristic-based Strategy [2, 23–26] relies on predefined rules or heuristics derived from metrics and source code models to detect code smells. History-based Strategy uses information about code evolution, such as changes in source code structure, to identify code smells [27, 28]. Optimization-based Strategy [29–31] employs optimization algorithms, such as genetic algorithms and evolutionary algorithms, often combined with metrics and code smell examples, to detect code smells. ML-based Strategy utilizes ML models like Decision Trees, Naïve Bayes, and Random Forest, etc. to train them on labeled datasets for code smell detection. Several ML models have been applied to detect a variety of code smells [32, 33, 42–45, 34–41].

Despite the extensive knowledge of various detected code smells and their relationships with other code smells (such as collocation, coupling, and inter-smell relations) there remains, to the best of our knowledge, no research that has utilized these insights for code smell detection using machine learning or similar techniques. In other words, no work has been done to classify the various code smell interactions and to use them in the detection process. One example of motivation for this study comes from the work of Lozano et al. [11], who identified a strong correlation between Feature Envy (FE) and Long Method (LM). They found that the presence of FE often signals a high likelihood of LM. Walter et al. [15] describe this as a plain support relationship, where FE, being easier to detect due to its simpler symptoms, may uncover broader underlying issues. This relationship suggests that FE and LM might share a common underlying problem, raising the question of whether detecting one could facilitate the detection of the other. Similarly, other significant concerns could also come up, such as how to leverage the knowledge of the Long Method code smell in Java to effectively detect and address the same issue in Python and other programming languages. These scenarios encourage us to explore further potential interactions between code smells. Therefore, understanding the various possible scenarios of code smell interactions is crucial for overall comprehension of code smells and further detection of code smells. Thus, this paper makes the following contributions:

1. *Identification of Code Smell Interaction Scenarios*: The paper introduces a set of identified scenarios for code smell interactions, with a focus on programming languages as domains. It also provides a detailed classification scheme to categorize these interactions.

2. *Development of a Novel Taxonomy*: The paper proposes a promising and new taxonomy for code smell interactions that integrates the identified scenarios and the classification scheme, offering a structured framework for understanding these interactions.

3. *Empirical Evaluation*: The paper presents experimental results from applying seven popular ML models on four distinct datasets, evaluating one of the identified interaction scenarios. This experimentation demonstrates the practical feasibility and effectiveness of the proposed taxonomy and classification scheme.

The organization of the paper is as follows. First, it presents the related work in section 2 along 2 key dimensions: code smell detection methods and code smell interactions. Second, section 3 illustrates the identified scenarios for code smell interactions. Subsequently, section 4 presents the proposed classification scheme for code smell interactions. It elaborates a new taxonomy for code smell interactions by merging the proposed classification scheme with identified scenarios for code smell interactions. This paper has handled one scenario called Inter smell detection within a domain (one of the identified scenarios for code smell interactions) within a set of different projects. For the same, the experimental methodology utilizing ML models is explained in section 5. Analysis and discussion of the results obtained are presented in section 6. The paper concludes by presenting final remarks in section 7.

## 2. Related work

This section discusses related work across two key dimensions: code smell interactions and code smell detection approaches. It is organized into two subsections: Subsection 2.1 focuses on related work concerning code smell interactions, while Subsection 2.2 reviews various approaches for code smell detection. Finally, the section concludes with a summary of key insights.

### 2.1 Code smell Interactions

According to a literature review, code smells do not occur in the code alone [5, 6] and is accompanied by other code smells [7–11]. Inter-smell relations happen when two or more code smells are connected or dependent on one another [46]. To identify the several relations among code smells, Pietrzak and Walter [15] investigated 830 classes in the Apache Tomcat 5.5.4 project and discovered six distinct relations between code smells: Data Class (DC) and Feature Envy (FE), Large Class(LC) for FE, Inappropriate Intimacy (II) and DC, and II and DC and FE. One industrial system and two open-source systems (Ebehandling, ElasticSearch (Version 1.2.1), and Apache Mahout (Version 0.7)) were examined by Yamashita et al. [47]. Using PCA with orthogonal rotation, they found inter-smell links for coupled and collocated smells. They found God Class (GC) and FE, as well as DC and FE, as significant smell relations.

Another study by Yamashita [48] used Principal Component Analysis (PCA) on one industrial system and two open-source software (Ebehandling, ElasticSearch, and Apache Mahout) to identify 15 different inter-smell patterns. Palomba et al. [9] evaluated the co-occurrences of 13 different code smells over 395 releases of 30 open-source software projects using association rule mining at the class level, and they found six pairs of code smells that frequently co-occurred: Complex Class (CC) and Message Chains (MC), Long Method (LM) and Long Parameter List (LPL), LM and FE, Spaghetti Code (SC) and LM, II and FE, Refused Bequest (RB) and Message Chains (MC). Lozano et al. [11] examined smell relations utilizing several releases of the three open-source projects: Log4j, Jmol, and JFreechart. This study discovered a correlation between three bad smells: FE, LM, and GC. An extensive study by Palomba et al. [13] on the co-occurrences of code smells employed 395 releases from 30 open-source software projects. They examined over 13 different code smells and found that six pairs of code smells regularly appeared. Walter et. al. [12] analysed various datasets which included all 92 systems in the latest version of Quality Corpus. They identified and experimentally confirmed common collocations of 14 code smells found in 92 Java systems using three methods: pairwise correlation analysis, principal component analysis (PCA), and associative rules. Thus, Java projects have been the main focus of the literature on inter-smell relations. Simultaneously, extensive research has been carried out in the field of code smell detection.

## 2.2 Code smell detection techniques

Recently, machine learning (ML)-based approaches have demonstrated significant success in code smell detection. The existing literature highlights substantial efforts to leverage ML techniques for identifying various code smells [17, 49]. The ML-based approach utilizes ML models like Decision Trees, Naïve Bayes, and Random Forest, etc. to train them on labeled datasets for code smell detection. Several ML models have been applied to detect a variety of code smells [32, 33, 42–45, 34–41]. It is noteworthy that we identified only two identical studies [50, 51] that explore the application of transfer learning for the detection of code smells. Both studies have explored the practicality of utilizing deep learning(DL) models and transfer learning (based on DL models for identifying four distinct code smells). The findings from the initial study [51] indicated that the efficacy of DL and transfer learning was contingent upon the specific type of smell, with transfer learning generally outperforming the associated DL models in terms of F1 score across most scenarios. The second subsequent study [50] indicated that the performance of DL models was contingent on specific conditions, and the efficacy of transfer learning was found to be on par with that of DL models. According to our proposed classification scheme of code smell, the technique of transfer learning used in papers [50, 51] falls as one of the subclasses of our proposed classification scheme.

Thus, the existing literature has explored code smell interactions in terms of collocation, coupling, and inter-smell relationships. However, to the best of our knowledge, no research has leveraged these insights to detect code smell through ML or similar techniques. In other words, no work has been done to classify the various code smell interactions and to use them in the detection process of further code smells. Therefore, understanding the various possible scenarios of code smell interactions is crucial for overall comprehension of code smells specifically for exploring their relationships and their detection.

## 3. Identified Scenarios for Code Smell Interactions

This section is divided into two sub-sections. Sub-section 2.1 outlines the problem formulation, detailing the structure of source and target data. Sub-section 2.2 defines key terms and introduces various scenarios and sub-scenarios for code smell interaction.

**3.1 Problem Formulation**

A code smell can depend on the domain [52–56] and the context [2]. The context could be a developer's viewpoint, a problem with the software's quality, or a problem with the design. A programming language, a particular application, or a setting can all be considered domains. The study focuses on domain-specific interactions specifically programming languages since each programming language has its distinct characteristics, such as syntax, semantics, idioms, and conventions, which can influence the appearance and detection of code smells. For example, code smells in Java may manifest differently from those in Python because of Java's object-oriented nature, static typing, and stricter class and method structure. Python, with its dynamic typing and flexibility in code design, may introduce different kinds of smells, making it challenging to use a single tool or approach to detect code smells across both languages. However, Python and Java share numerous common code smells such as Long method, Large class, Duplicate code, etc. A tool or approach that is effective in detecting Long method smells in Java may not be applicable or accurate for detecting Long method smells in Python. This is because the syntax and semantics of each programming language shape how code smells are expressed. Java is more verbose due to explicit type declarations, access modifiers, and syntax requirements like semicolons and braces. This verbosity can make Java methods appear longer compared to Python methods with the same logic. Python's syntax is more concise and readable. However, this can sometimes mask the underlying complexity, and a long method in Python may appear shorter due to fewer lines of code. As a result, it is important to adopt domain-specific approaches to detect and address code smells effectively. Therefore, considering language as a domain, the problem of code smell interaction has been formulated as follows:

Let's define the data involved:

- Source Data ($X_s$, $Y_s$)
  - $X_s = \{x_1, x_2 \ldots, x_p\}$ represent the source feature space of dimension p, where each $x_i$ is a feature in the source data. A feature can represent a source code metric or another characteristic of the source code such as source code embedding[57–59], with a specific set of features depending on the particular code smell being analyzed.
  - $Y_s = \{y_1, y_2 \ldots, y_n\}$ represent the source label set, where each $y_i \in \{0,1\}$ indicates the presence (1) or absence (0) of code smell $CS_1$ for the corresponding instance in $X_s$.
- Target data ($X_t$, $Y_t$):

- $X_t = \{x_{1'}, x_{2'}, \ldots, x_{s'}\}$ represent the target feature space of dimension s, where each $x_{i'}$ is a feature in the target data.
- $Y_t = \{y_{1'}, y_{2'}, \ldots, y_{m'}\}$ represent the target label set, where each $y_{i'} \in \{0,1\}$ indicates the presence (1) or absence (0) of code smell $CS_2$ for the corresponding instance in $X_t$.

Assume that Source data and target data belong to different programming languages. The goal is to predict the target labels $Y_t$ for target instances in $X_t$, using the knowledge transferred from the source data $(X_s, Y_s)$. An ML technique such as transfer learning or other can be used to improve the learning from one domain by transferring the knowledge from the same or related domain to the other. So, source data $[X_s, Y_s]$ can be used to find target labels in $Y_t$. Since source data and target data may come from different developers and might contain different feature spaces for different types of code smells. So, a scenario may be identified based on source and target data parameters.

### 3.2 Key Terms & Identified Scenarios and Sub-scenarios

Consider the source and target data as given in Table 1 and definitions of source and target data in sub-section 3.1 above. The table displays four parameters for source data and target data. The parameters are features ($X_s$ and $X_t$), data labels ($Y_s$ and $Y_t$), the type of code smell ($CS_1$ and $CS_2$), and code language ($L_1$ and $L_2$). The interplay of features, code smells, and languages leads to multiple scenarios. Table 2 outlines three distinct scenarios categorized based on the type of code smell (intra-smell or inter-smell), the nature of the source and target data features (homogeneous or heterogeneous), and the domain involved (within-domain or cross-domain). Each scenario is named according to these key factors. The abbreviations used are **c** for cross-domain, **h** for heterogeneous, **$D_t$** for detection, **D** for domain, and **S** for smell. This structure helps in understanding the variations in detection approaches across different data and domain settings. Here's a description of the key terms used in the Table 1 and Table 2:

**Table 1** Notations

| Parameters | Feature space | Code smell | Language | Labels |
|---|---|---|---|---|
| Source Data | $X_s$ | $CS_1$ | $L_1$ | $Y_s$ |
| Target data | $X_t$ | $CS_2$ | $L_2$ | $Y_t$ |

1. Code Smell ($CS_1$, $CS_2$): Code Smell (CS) refers to indicators in the source code that might suggest deeper problems in the system design, leading to potential maintenance issues in the future. Each smell type (e.g., $CS_1$, $CS_2$) represents a specific kind of code smell, such as "Large Class," "Long Method," etc.
2. Code Language (L1, L2):
   - $L_1$ (Source Language): The programming language used in the source dataset. For example, Java or Python.
   - $L_2$ (Target Language): The programming language used in the target dataset. This can be the same as or different from $L_1$.
3. Intra-Smell vs. Inter-Smell:
   - Intra-Smell: The same type of code smell is being detected in both the source and target datasets (e.g., $CS_1$ in both cases).
   - Inter-Smell: Different types of code smells are being detected between the source and target datasets (e.g., $CS_1$ in the source, $CS_2$ in the target).
4. Homogeneous Features vs. Heterogeneous Features:
   - Homogeneous Features: When the features ($X_s$ and $X_t$) in the source and target datasets are of the same type, meaning that the same metrics are used for both.
   - Heterogeneous Features: When the features in the source and target datasets differ, indicating that different metrics are used to represent the source and target codebases.
5. Within Domain vs. Cross-Domain (Cross-language):
   - Within Domain: The source and target data come from the same programming domain or context, meaning they share similar environments (e.g., both datasets are from the same programming language).
   - Cross-Domain (Cross - language): The source and target data come from different domains,

such as different programming languages ($L_1 \neq L_2$), indicating a transfer learning or domain adaptation scenario.
6. Scenario: A scenario defines the broad context in which code smells are being detected. It outlines the major conditions, such as Domain and Features.
7. Sub-scenario: Each sub-scenario defines a particular combination of the following:
    a) Features ($X_s$ and $X_t$): Whether the source and target features are the same (homogeneous) or different (heterogeneous).
    b) Code Smell Types ($CS_1$ and $CS_2$): Whether the source and target code smells are the same (intra-smell) or different (inter-smell).
    c) Programming Languages ($L_1$ and $L_2$): Whether the source and target datasets come from the same domain (same programming language) or from different domains (cross-domain, where languages differ).

Table 2 outlines three scenarios and their respective sub-scenarios. Examples for each sub-scenario are provided in Table A.1 in Appendix A to facilitate understanding.

**Table 2** Identified scenarios

| Sub-scenario | Rule | Abbreviation used |
|---|---|---|
| *Within Domain smell detection* | | |
| *Scenario 1. Within Domain smell detection (Homogeneous)* | | |
| 1.1 | If $X_s = X_t$ and $CS_1 = CS_2$ and $L_1 = L_2$ | Intra-smell Detection within domain (Intra $SD_tD$) |
| 1.2 | If $X_s = X_t$ and $CS_1 \neq CS_2$ and $L_1 = L_2$ | Inter-smell Detection within domain (Inter $SD_tD$) |
| *Scenario 2. Within Domain smell detection (Heterogeneous)* | | |
| 2.1 | If $X_s \neq X_t$ and $CS_1 = CS_2$ and $L_1 = L_2$ | Intra smell Detection within heterogeneous domain (Intra $SD_thD$) |
| 2.2 | If $X_s \neq X_t$ and $CS_1 \neq CS_2$ and $L_1 = L_2$ | Inter smell Detection within heterogeneous domain (Inter $SD_thD$) |
| *Scenario 3. Cross-domain (Cross - language) smell detection* | | |
| 3.1 | If $X_s = X_t$ and $CS_1 = CS_2$ and $L_1 \neq L_2$ | Intra smell Detection within cross-domain (Intra $SD_tcD$) |
| 3.2 | If $X_s = X_t$ and $CS_1 \neq CS_2$ and $L_1 \neq L_2$ | Inter smell Detection within cross-domain detection (Inter $SD_tcD$) |
| 3.3 | If $X_s \neq X_t$ and $CS_1 = CS_2$ and $L_1 \neq L_2$ | Intra smell Detection within heterogeneous cross-domain (Intra $SD_thcD$) |
| 3.4 | If $X_s \neq X_t$ and $CS_1 \neq CS_2$ and $L_1 \neq L_2$ | Inter smell Detection within heterogeneous cross domain (Inter $SD_thcD$) |

Sub-scenario 1.1 of Scenario 1, termed Intra Smell Detection within Domain (Intra $SD_tD$), involves detection within the same language domain. This means that both the source and target data share the same set of features, labels, code smells, and language, though they may have similar or differing distributions. Several studies [32, 33, 42–45, 34–41] utilizing ML models for code smell detection fall under Sub-scenario 1.1. In cases where the source and target data distributions are similar, conventional ML models are applicable. However, when the distributions differ, a homogeneous transfer learning [60] approach becomes more appropriate for this scenario.

Sub-scenario 1.2 of Scenario 2 focuses on inter-smell detection, which involves identifying different code smells present in both the source and target datasets. In the case of Inter Smell Detection within Domain (Inter $SD_tD$), there are two datasets for two distinct code smells, but both datasets share the same set of features. Transfer learning techniques can be applied here: a model trained on one dataset can be used to predict related code smells in the other dataset. Similar approaches apply to other sub-scenarios. A recent study by Gupta et al. [61] can be classified under Sub-scenario 2.1, referred to as Intra-smell Detection within a Heterogeneous Domain (Intra $SD_thD$). In this study, two code smells—Temporary Field and Long Method—were detected using a customized heterogeneous transfer learning technique. The authors assumed heterogeneous feature spaces between source and target data, with the Java programming language as the domain for analysis.

When the source and target datasets differ in language, this is termed Cross-Domain Smell Detection, further categorized into four sub-scenarios (3.1 to 3.4) in Scenario 3. Transfer learning techniques [60] are also useful in these cases. For Intra Smell Detection within Cross-Domain (Intra $SD_tcD$), there are two datasets for the same code smell but in different languages. One dataset (source) is used to train a (let us say transfer) learning model, while the other dataset (target) is used to test the model's prediction capabilities. Similar interpretations apply to other sub-scenarios (3.3 and 3.4).

Studies [50, 51] fall under Sub-scenario 3.3 referred to as Intra-smell Detection within a Heterogeneous Cross-Domain (Intra $SD_thcD$). It is because these studies focused on detecting code smells across different domains (Java and C#) that have heterogeneous feature spaces. Both studies employed deep learning(DL) models to facilitate cross-domain knowledge transfer, where DL models trained on Java projects were tested on C# projects, and vice versa.

## 4. Proposed Classification Scheme and taxonomy for Code Smell Interactions

This section introduces the proposed classification scheme for code smell interactions and elaborates on the corresponding taxonomy, which integrates the classification scheme with the identified scenarios discussed in Section 3.

Figure 1 illustrates a novel classification scheme for code smell interactions, utilizing programming language as the domain of focus. Projects are classified based on their alignment within a single domain or across multiple domains. Smell interactions can manifest within a single project, across various projects, or among a collection of projects. When these projects encompass various domains, two scenarios emerge: (1) interactions might exhibit the same code smell across domains, or (2) interactions could involve different code smells between domains. For instance, take an example of two projects—one developed using Java and the other utilizing C#. Identifying Long method smells in a Java project can provide valuable insights for predicting similar issues in a C# project, as code smells frequently indicate underlying structural issues across different programming languages.

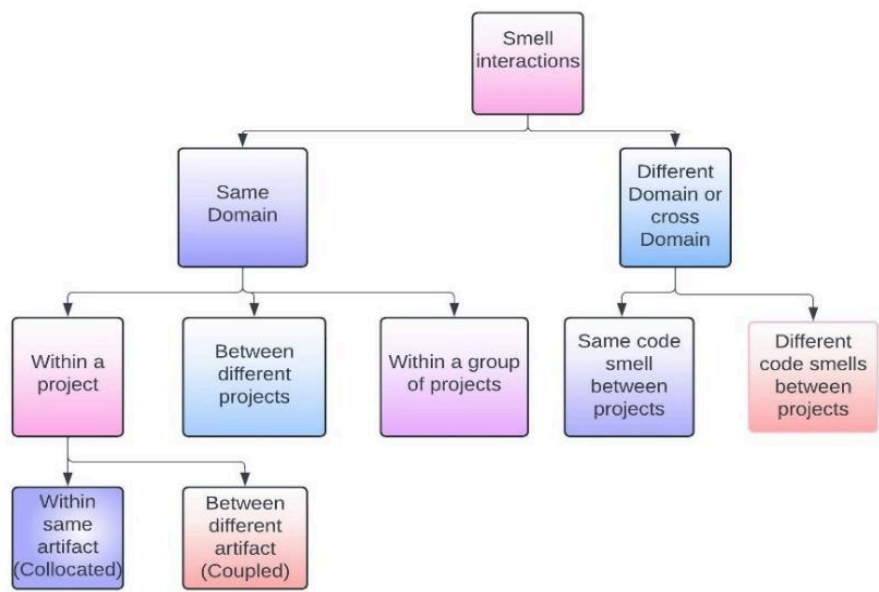

**Fig.1** Proposed classification scheme for smell interactions

Building on this classification scheme, Figure 2 presents a taxonomy for code smell interactions by integrating the proposed classification with various identified scenarios. The taxonomy organizes smell

interactions into specific subclasses:

Figure 2 illustrates a novel taxonomy for code smell interactions by integrating the proposed classification scheme with the identified scenarios. It demonstrates that sub-scenarios 1.1 and 1.2 fall under the subclass of interactions occurring between different projects or within a group of projects under the same domain category, as both belong to the same programming language domain. Within the same domain category, Inter $SD_tD$ and Inter $SD_thD$ correspond to collocated code smells. Similarly, sub-scenarios 2.1 and 2.2 are also classified under the same subclass, representing interactions within the same domain across different projects. On the other hand, sub-scenarios 3.1 and 3.3 are categorized under the subclass of same code smell interactions across projects in the cross-domain category, reflecting scenarios where the same smell occurs across different domains (languages). Finally, sub-scenarios 3.2 and 3.4 fall under the subclass of different code smell interactions across projects, also within the cross-domain category, where different smells interact across projects from different domains. This taxonomy unifies various scenarios of smell interactions based on domain and smell types.

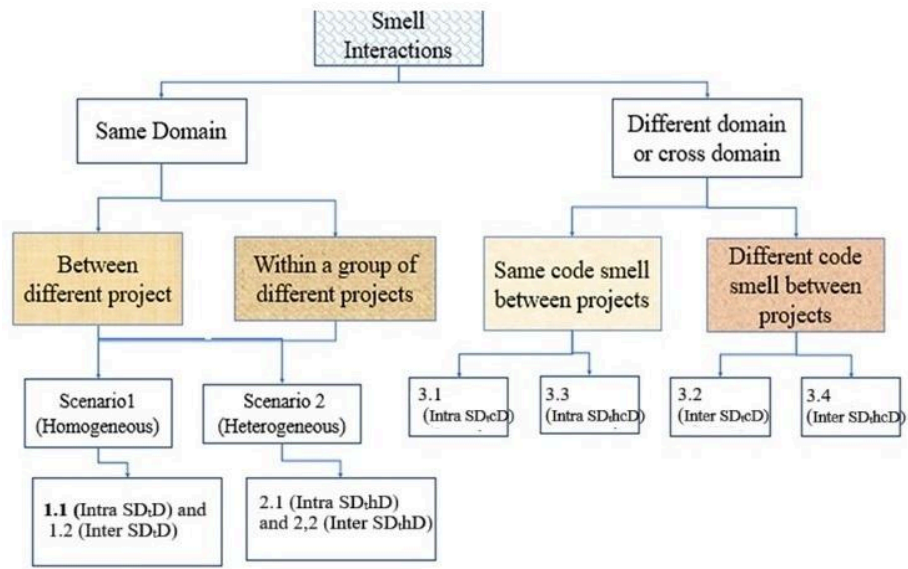

**Fig. 2** Taxonomy for code smell interactions

## 5. Inter $SD_tD$ using ML

This section outlines the experimental methodology and datasets used for sub-scenario 1.2, referred to as Inter-smell Detection within Domain (Inter $SD_tD$), which falls under the category of interactions across different projects within the same domain in the proposed taxonomy for code smell interactions. Sub-section 5.1 introduces the selected publicly available code smell datasets. Sub-section 5.2 details the methodology employed for the experiment. Finally, Sub-section 5.3 discusses the various performance metrics used to evaluate Inter $SD_tD$.

**5.1 Datasets used**

The literature has been surveyed to identify publicly available datasets appropriate for the experiment. After exploration of the literature only 6 research papers [12], [33], [62], [50], [63], [64] were found that provided their code smell datasets in the public domain. For analysis, all the mentioned datasets were manually downloaded. The downloaded dataset and the related paper were simultaneously examined for the data contained in them.
The paper has used 4 published datasets by Arcelli Fontana et al. [33] for 4 different code smells: long method, feature envy, data class, and god class because they were the only publicly accessible datasets with several metrics and types of smells. Hence they were judged appropriate for the experimentation using ML algorithms.

To construct the dataset, the authors [33] have chosen 74 diverse Java open-source systems from the Qualitas Corpus [65], representing a range of application domains. They computed various project metrics at both the class and method levels using the *DFMC4J (Design Features and Metrics for Java)* tool. Code smells were then identified and labeled using a combination of tools, including *iPlasma, PMD1, Fluid Tool, AntiPattern Scanner*, and specific rules. Following this, three trained students manually validated the results. Each student performed individual evaluations of code smells before collaborating to reach a consensus, which led to the development of guidelines for identifying each reported code smell (Table 3 outlines the four different code smells along with their associated guidelines for their detection). As a result, four unique datasets were generated, with each dataset aligned to a specific code smell: long method, feature envy, data class, and God class. The datasets were structured with an increased number of features at the method level, reflecting the differences in granularities. The appendix of the paper [33] contains an extensive enumeration of features along with their definitions. The dataset was organized to consist of one-third smelly samples and two-thirds non-smelly samples for each code smell.

**Table 3** Code smells and corresponding rules for detection[33]

| Code smell | Rules / Guidelines |
|---|---|
| God class | o They contain extensive and complex methods.<br>o They are huge in size.<br>o They usually expose numerous methods.<br>o They have access to several attributes from a wide range of classes. |
| Data class | o They do not contain complex methods.<br>o They can make accessible a limited number of non-accessor simple methods.<br>o Their attributes should be defined as either public or made accessible via accessor methods. |
| Long method | o They exhibit a high level of complexity.<br>o They typically access numerous attributes, a significant number of class members, and a large number of used variables, including those from ancestors.<br>o They are massive.<br>o They usually contain several parameters. |
| Feature envy | o They bear access to a greater range of foreign attributes compared to local ones.<br>o They mainly utilize the attributes of some foreign classes. |

The following steps were carried out to prepare the datasets for the experiment and to identify the cases of Inter SDDt.

1. The published datasets [33] were downloaded and analyzed to examine the features they included, along with other statistical characteristics such as the range of feature values and the total number of samples. Table 4 details that each dataset consisted of one-third smelly samples (+ve) and two-thirds non-smelly samples (-ve). Additionally, Table 5 provides information on the number and percentage of missing values, as well as the names of features with missing values for each dataset. Missing values were addressed using the mean value imputation method, a widely recognized and straightforward approach in the literature.
2. The datasets for long method and feature envy had identical features, while the datasets for God class and data class also shared the same features. Based on this analysis, four distinct cases were identified for experimentation (using ML) for sub-scenario 1.2, referred to as Inter $SD_tD$. These four cases are as follows.

    (a) from Long Method to Feature Envy (LM to FE) denoted by $LM \rightarrow FE$
    (b) from Feature Envy to Long Method (FE to LM) denoted by $FE \rightarrow LM$
    (c) from God Class to Data Class (GC to DC) denoted by $GC \rightarrow DC$
    (d) from Data Class to God Class (DC to GC) denoted by $DC \rightarrow GC$

The study does not consider cases such as LM→GC or LM→DC, etc. as these fall outside the scope of *Inter Smell Detection within the Same Domain* (Inter $SD_tD$). Instead, they belong to sub-scenario 2.2, referred to as *Inter Smell Detection within Heterogeneous Domains* (Inter $SD_thD$). Traditional machine learning techniques do not apply to Inter $SD_thD$ due to domain heterogeneity. In such scenarios, heterogeneous transfer learning[66] would be more suitable. Since the current study focuses exclusively on Inter $SD_tD$, these cases have been excluded from consideration. When the source data contains multiple code smells and the target data involves a single code smell (e.g., training a model on Feature Envy and Long Method to detect God Class), a heterogeneous transfer learning technique can be applied.

**Table 4** Code smells Datasets and their missing values

| SI No. | Dataset name | Number of features | No. of smelly samples | No. of –non-smelly samples | Count of missing values | Missing values % | Feature - count of missing values |
|---|---|---|---|---|---|---|---|
| 1 | DC | 62 | 140 | 280 | 75 | 0.0028% | NMO-19, NIM -19, NOC -9, WOC -28 |
| 2 | GC | | | | 76 | 0.0029% | NMO-20, NIM-20, NOC-8, WOC -28 |
| 3 | LM | 82 | | | 92 | 0.0026% | NOC-3, WOC-31 NIM-29, NMO-29 |
| 4 | FE | | | | 92 | 0.0026% | NOC-3, WOC-31 NIM-29, NMO-29 |

## 5.2 Methodology of Experiment

Machine learning (ML) methodologies require the utilization of training and testing datasets tailored to a particular problem domain. In ML, the datasets used for training and testing frequently originate from one source and encompass an identical array of features. In the process of employing ML to detect code smells, a dataset is partitioned into training and testing subsets. An ML model is then developed using the training subset and subsequently assessed using the testing subset. Performance metrics such as the F-measure, accuracy, precision, and recall are commonly utilized to evaluate the effectiveness of ML models. Consequently, the training and testing datasets exhibit identical code smells. The analysis presented in Figure 4 indicates that our approach utilized two distinct datasets corresponding to two separate code smells within our ML model for Inter $SD_tD$. A single dataset is utilized for the training of an ML model, while a separate dataset is employed to evaluate the performance of the trained model. Consequently, with a different dataset, the detection of code smell is anticipated. The performance of the ML model is subsequently assessed. Therefore, rather than utilizing a single dataset as is standard practice, Inter $SD_tD$ leveraging ML necessitates the use of two datasets. Figure 3 illustrates the method employed in the study.

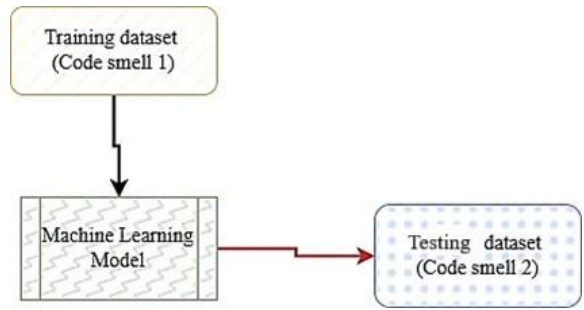

**Fig. 3** ML for inter-smell detection

Seven well-established ML models [17] (Random Forest, Decision Tree, Logistic Regression, Naive Bayes, Multilayer Perceptron, Support Vector Machine, K-nearest neighbour Classifier) were implemented across four datasets for four distinct cases of Inter $SD_tD$. The investigation of these four cases was conducted under two specific conditions: (1) Balanced source data, characterized by a ratio of positive (smelly) to negative (non-smelly) samples of 0.33, and (2) Unbalanced source data, with a ratio of positive to negative samples at 0.10. This distinction is crucial, as imbalanced source data can significantly influence the performance of an ML model.

Thus, *LM→FE* refers to the process of employing an ML model that is trained on the LM dataset to predict labels for the FE dataset. As a result, the FE dataset acts as the testing dataset, and the LM dataset as the training dataset. Thus, the purpose of employing ML in the Inter $SD_tD$ is to investigate the feasibility of detection of FE code smell by utilizing knowledge of LM code smell. The same interpretation applies to all other cases of Inter $SD_tD$.

The performance of an ML model is significantly affected by the values of its hyperparameters. Therefore, it is essential to determine the ideal values of these hyperparameters to achieve optimal performance. Hyperparameters refer to the parameters that are established before the commencement of the learning process. The learning process is governed by them, instead of deriving knowledge from the data. Consequently, the hyperparameters of the analysed ML models have been adjusted to achieve their peak performance, following the guidelines provided in the study conducted by Yang and Shami [67]. The optimization process varies for different ML models due to their specific hyperparameters. Below is a summary of the key hyperparameters for each model:

a) In the K-Nearest Neighbors (KNN) algorithm, the value of k, denoting the number of nearest neighbors to consider, is an essential hyper-parameter.
b) The selection of the kernel function is a critical hyperparameter in the support vector machine algorithm.
c) The smoothing parameter α is a crucial hyperparameter in the Naïve Bayes algorithm.
d) The decision tree algorithm includes several hyperparameters that can be adjusted for optimal performance. These hyperparameters include the splitting criteria, the maximum number of features to consider when generating the best split, and the maximum number of leaf nodes allowed. The minimum number of data points required to split a decision node is represented by the parameter 'min_samples_split', while the minimum number of data points required to obtain a leaf node is represented by the parameter 'min_samples_leaf'.
e) The Random Forest (RF) algorithm employs the bagging technique to aggregate multiple decision trees. In the Random Forest (RF) algorithm, Decision Trees (DTs) are constructed using randomly generated subsets of the data. The final classification result is determined by selecting the class with the highest voting count among the DTs. The Random Forest (RF) algorithm requires tuning of a crucial hyperparameter, namely the number of decision trees to be aggregated. The hyperparameter 'n_estimators' is the designated term used in the sklearn library for the same.
f) The logistic regression model in the sklearn library of Python includes hyperparameters such as penalty (which determines the penalty for a regularization method), C (which controls the regularization strength), and solver (which specifies the algorithm used for optimization).
g) However, they did not specify the hyperparameters for the multilayer perceptron algorithm.

The experiment was carried out on an independent personal computer system using Python 3[68] via Keras. The GridSearchCV[69] function was utilized to optimize the hyperparameters across different ML models, aiming to achieve their best performance. The GridSearchCV function conducts a comprehensive search across a defined grid of parameter values, as specified by the param_grid parameter, for the designated parameters of an ML algorithm. Appendix A, specifically Table A.2, presents the values of the hyperparameters utilized across different ML models. It also provides the names of the modules used in the sklearn library, and the values utilized in the GridSearchCV function. For each of the four cases, the same hyperparameter values were applied under both conditions (balanced and unbalanced source data). The sklearn is a popular Python library used for ML and data analysis. This library contains a wide range of tools for building ML models, including algorithms for classification, regression, clustering, dimensionality reduction, and preprocessing of data.

**5.3 Performance Metrics for Inter $SD_tD$**

The paper [70] states that performance metrics or measures may be chosen according to the requirement for a given problem. The key performance metrics in our experiment are precision and Area Under the Curve (AUC) [71] as the objective is to detect the presence of a code smell using the available knowledge of the presence of another code smell. The precision metric indicates the ability of an ML model to identify the existence of a code smell based on the knowledge of another code smell. When comparing ML models to determine which one is more robust, the Area Under the Curve (AUC) metric has been used. Thus, the experiment has used these performance measures. Recent articles [72] [71] have demonstrated that AUC is a superior performance statistic compared to others. Below is a concise discussion of various performance measures within the context of this paper.

1. PRECISION (P)

Precision refers to the percentage of correctly predicted code smells. A higher precision signifies a superior model, with the maximum achievable value being 100%. High precision is desirable because it means the model is accurately identifying code smells without producing many false positives.

2. AREA UNDER AN ROC CURVE (AUC)

The AUC (Area Under the Curve) is a measure of the model's ability to differentiate between classes at various threshold values. It reflects the degree of separability of classes. The AUC value varies between 0 and 1, where a higher AUC value signifies a more effective model performance. The ML model performs best when the AUC is closer to 1, which indicates that there is a lot of separability. AUC around 0 denotes the lowest degree of separability and is indicative of a poor model. An AUC of 0.5 indicates that the ML model is unable to distinguish between various classes.

# 6. Results and discussion

This section presents the results under 3 sub-sections. Sub-section 6.1 discusses the results of Inter $SD_tD$ on balanced source data. Sub-section 6.2 presents and discusses the results on unbalanced source data. The last sub-section 6.3 compares the results of sub-sections 6.1 and 6.2. In the end, it analyzes the various codes belonging to 4 different code smells (long method, feature envy, god class, and data class) which have been extracted from the experimental projects of the downloaded datasets [33] to further confirm the results.

## 6.1 Results of Inter $SD_tD$ on balanced data

The experiment's results in a situation of balanced data are displayed in Table 5. For $LM{\rightarrow}FE$, it is observed that the k-nearest neighbor is best in terms of precision with a score of 0.75. However, it is not the best in terms of AUC since its AUC score is 0.56. In other words, it can be said that the k-nearest neighbor detects the presence of feature envy code smell from the knowledge of the presence of long method code smell. In the same way, all other cases can be interpreted. For the $FE{\rightarrow}LM$, it is observed that again k-nearest neighbor is best in terms of precision with a score of 0.66. However, it is not the best in terms of AUC since its AUC score is 0.56. But Naïve Bayes can be assumed to be overall better than others in this case since it has a precision of 0.84 (close to that of k-nearest neighbor) and AUC of 0.78. Thus, it can be concluded that the presence of the long method detects the presence of feature envy and vice versa. This shows the compatibility with the information available in the literature. It was previously discovered by [9, 11, 12] that feature envy and long method code smells co-occur.

**Table 5** Results for Sub-scenario1.2 (Inter $SD_tD$) on Balanced data

| SI No. | ML model/ Code smell | $LM{\rightarrow}FE$ | | $FE{\rightarrow}LM$ | | $GC{\rightarrow}DC$ | | $DC{\rightarrow}GC$ | |
|---|---|---|---|---|---|---|---|---|---|
| | | Precision | AUC | Precision | AUC | Precision | AUC | Precision | AUC |
| 1 | Random Forest | 0.55 | 0.74 | 0.63 | 0.74 | 0.00 | 0.29 | 0.00 | 0.29 |
| 2 | Decision Tree (C4.5) | 0.55 | 0.74 | 0.56 | 0.67 | 0.00 | 0.30 | 0.00 | 0.29 |
| 3 | Logistic Regression | 0.50 | 0.73 | 0.61 | 0.73 | 0.00 | 0.23 | 0.00 | 0.24 |
| 4 | Naïve Bayes | 0.55 | 0.74 | 0.84 | 0.78 | 0.00 | 0.31 | 0.02 | 0.05 |
| 5 | Multilayer Perceptron | 0.54 | 0.69 | 0.50 | 0.51 | 0.18 | 0.36 | 0.04 | 0.14 |
| 6 | K-nearest neighbor | 0.75 | 0.56 | 0.90 | 0.66 | 0.18 | 0.40 | 0.33 | 0.50 |
| 7 | Support Vector Machine (SVM) | 0.55 | 0.74 | 0.64 | 0.73 | 0.00 | 0.29 | 0.00 | 0.30 |

For the cases: $GC{\rightarrow}DC$ and $DC{\rightarrow}GC$, precision is very poor (close to zero) in almost all ML models. Due to a low AUC (<0.5), the ML models failed to distinguish between smelly and non-smelly samples and performed worse than random guessing. Thus, due to poor precision and poor AUC, it can be concluded that the presence of the God class does not detect the presence of the data class and vice versa i.e., the God class and data class do not co-occur in the projects. Thus, the results are compatible with the existing facts available in the literature that depict that there is a utilized relationship between the God Class and Data Class and they do not co-occur Fontana et al. [7].

## 6.2 Results of Inter Inter $SD_tD$ on unbalanced data

The experiment's results in a situation of unbalanced data are displayed in Table 6. For $LM{\rightarrow}FE$, it is observed that SVM is best in terms of precision with a score of 0.71. However, it is not the best in terms of AUC since its AUC score is 0.69. However, SVM is considered a better option than others since it has quite good scores in both precision and AUC. In other words, it can be said that SVM can detect the presence of Feature Envy code smell from the knowledge of the presence of Long Method code smell. For $FE{\rightarrow}LM$, it is observed that again SVM is best in terms of precision and AUC. Thus, it can be concluded that the presence of a Long Method detects the presence of Feature Envy and vice versa utilizing SVM.

**Table 6** Results for sub-scenario1.2 (Inter SD$_t$D ) on unbalanced data

| SI No. | ML model/ Code smell | LM→FE | | FE→LM | | GC→DC | | DC→GC | |
|---|---|---|---|---|---|---|---|---|---|
| | | Precision | AUC | Precision | AUC | Precision | AUC | Precision | AUC |
| 1 | Random Forest | 0.55 | 0.74 | 0.69 | 0.67 | 0.00 | 0.33 | 0.00 | 0.35 |
| 2 | Decision Tree (C4.5) | 0.53 | 0.73 | 0.59 | 0.63 | 0.00 | 0.30 | 0.00 | 0.37 |
| 3 | Logistic Regression | 0.56 | 0.64 | 0.69 | 0.67 | 0.00 | 0.41 | 0.00 | 0.38 |
| 4 | Naïve Bayes | 0.55 | 0.74 | 0.00 | 0.50 | 0.00 | 0.31 | 0.00 | 0.50 |
| 5 | Multilayer Perceptron | 0.00 | 0.50 | 0.00 | 0.50 | 0.00 | 0.49 | 0.00 | 0.47 |
| 6 | K-nearest neighbor | 0.62 | 0.63 | 0.67 | 0.53 | 0.00 | 0.50 | 0.00 | 0.47 |
| 7 | Support Vector Machine | 0.71 | 0.69 | 0.71 | 0.69 | 0.00 | 0.37 | 0.00 | 0.33 |

This shows the compatibility with the information available in the literature as already discussed in the above sub-section 6.1. For the cases: $GC→DC$ and $DC→GC$, the precision is consistently zero across all ML models. AUC score is also very low. Thus, it can be concluded that the presence of a God Class does not detect the presence of a Data Class and vice versa, meaning that the God Class and Data Class do not co-occur in the projects. Thus, the results are consistent with the existing information available in the literature as outlined in section 6.1

### 6.3 Comparison of results of Inter SD$_t$D on balanced and unbalanced data

Table 7 shows the comparison of results for balanced and unbalanced data. For the long method to feature envy, the k-nearest neighbor is best for balanced data in terms of precision, but SVM is best in terms of precision (0.71) for unbalanced data. It is also observed that the performance of some models such as MLP changed drastically between balanced and unbalanced data.

For $FE→LM$, it is observed that the AUC score is less in a situation of unbalanced data than compared to balanced data. Naive Bayes and MLP performed very poorly in the situation of unbalanced data compared to balanced data. Also, Naïve Bayes performed the best overall in a situation of balanced data, but SVM was the overall winner in a situation of unbalanced data.

For the cases: $GC→DC$ and $DC→GC$, in the situation of unbalanced data, precision is zero in all ML models. While, in the situation of balanced data, precision is close to zero in some ML models. Also, it can be observed that the performance of ML models in all 4 cases deteriorates somewhat.

Thus, it can be concluded that in both balanced and unbalanced data scenarios, the results indicate that the presence of a Long Method (LM) can be used to detect Feature Envy (FE) and vice versa, suggesting that these code smells tend to co-occur. Conversely, it is inferred that in both situations, the presence of a God Class (GC) cannot reliably detect the presence of a Data Class (DC), indicating that these two code smells do not co-occur. This finding aligns with the information available in the literature, as discussed in earlier sections. The failure of ML models to accurately detect the GC and DC code smells is likely due to the significant differences in their data distributions, which greatly deteriorate the model's performance. This was confirmed by the Maximum Mean Discrepancy (MMD) test which revealed a substantial variation in the marginal probability distribution (MPD between the datasets (LM vs. FE and DC vs. GC). The MMD test serves as a technique for identifying differences between two distributions by analysing the expected values of functions within a characteristic Reproducing Kernel Hilbert Space (RKHS) [73]. It has been thoroughly and efficiently utilized in transfer learning applications to ensure similarity in the MPD between training and test data [74, 75].

**Table 7** Comparison of results for balanced and unbalanced data

| Machine Algorithm/ Code smell | Source data | Performance measure | Random Forest | Decision Tree (C4.5) | Logistic Regression | Naïve Bayes | Multilayer Perceptron | K-nearest neighbour | Support Vector Machine |
|---|---|---|---|---|---|---|---|---|---|
| LM →FE | Balanced | Precision | 0.55 | 0.55 | 0.50 | 0.55 | 0.54 | 0.75 | 0.55 |
|  |  | AUC | 0.74 | 0.74 | 0.73 | 0.74 | 0.69 | 0.56 | 0.74 |
|  | Unbalanced | Precision | 0.55 | 0.53 | 0.56 | 0.55 | 0.00 | 0.62 | 0.71 |
|  |  | AUC | 0.74 | 0.73 | 0.64 | 0.74 | 0.50 | 0.63 | 0.69 |
| FE→LM | Balanced | Precision | 0.63 | 0.56 | 0.61 | 0.84 | 0.50 | 0.90 | 0.64 |
|  |  | AUC | 0.74 | 0.67 | 0.73 | 0.78 | 0.51 | 0.66 | 0.73 |
|  | Unbalanced | Precision | 0.69 | 0.59 | 0.69 | 0.00 | 0.00 | 0.67 | 0.71 |
|  |  | AUC | 0.67 | 0.63 | 0.67 | 0.50 | 0.50 | 0.53 | 0.69 |
| GC→DC | Balanced | Precision | 0.00 | 0.00 | 0.00 | 0.00 | 0.18 | 0.18 | 0.00 |
|  |  | AUC | 0.29 | 0.30 | 0.23 | 0.31 | 0.36 | 0.40 | 0.29 |
|  | Unbalanced | Precision | 0.00 | 0.00 | 0.00 | 0.00 | 0.00 | 0.00 | 0.00 |
|  |  | AUC | 0.33 | 0.30 | 0.41 | 0.31 | 0.49 | 0.50 | 0.37 |
| DC→GC | Balanced | Precision | 0.00 | 0.00 | 0.00 | 0.02 | 0.04 | 0.33 | 0.00 |
|  |  | AUC | 0.29 | 0.29 | 0.24 | 0.05 | 0.14 | 0.50 | 0.30 |
|  | Unbalanced | Precision | 0.00 | 0.00 | 0.00 | 0.00 | 0.00 | 0.00 | 0.00 |
|  |  | AUC | 0.35 | 0.37 | 0.38 | 0.50 | 0.47 | 0.47 | 0.33 |

To further validate the findings, an analysis of the extracted source codes from the experimental projects was conducted. For instance, a few examples have been presented along with their analysis in the following sections. The source codes for the four identified code smells are provided below, with the detection rules applied to these code smells displayed in Table 3 in sub-section 5.1. The source code information was obtained from the downloaded datasets [33], with all source codes extracted from the original repository of Qualitas Corpus. Unnecessary comments were removed from the code samples, as they were not relevant to the analysis.

*Data class*:

The class shown below in Figure 4 was extracted from HTMLScanner.java which is part of the nekohtml-1.9.14 project, one of the 74 Java open-source projects from Qualitas Corpus. The code in Figure 5 represents a Java class named LocationItem that implements the HTMLEventInfo interface and provides methods to retrieve and set location information for HTML events.

The class is declared as protected static, which means it can only be accessed within its package or subclasses. It implements the HTMLEventInfo interface and also implements the Cloneable interface. The class contains several integer fields (fBeginLineNumber, fBeginColumnNumber, fBeginCharacterOffset, fEndLineNumber, fEndColumnNumber, EndCharacterOffset) to store location information. It has another constructor that takes another LocationItem object as a parameter and copies its values to initialize a new LocationItem object. It has several public methods (setValues(): Sets the values of the location item. getBeginLineNumber(), etBeginColumnNumber(), getBeginCharacterOffset(), getEndLineNumber(), getEndColumnNumber(), etEndCharacterOffset(): Retrieve the corresponding location information.isSynthesized(): Returns false indicating that the event is not synthesized.).

```
protected static class LocationItem implements HTMLEventInfo, Cloneable {
protected int fBeginLineNumber;
protected int fBeginColumnNumber;
protected int fBeginCharacterOffset;
protected int fEndLineNumber;
protected int fEndColumnNumber;
protected int fEndCharacterOffset;
public LocationItem() { }
LocationItem(final LocationItem other) {
setValues(other.fBeginLineNumber, other.fBeginColumnNumber, other.fBeginCharacterOffset,
other.fEndLineNumber, other.fEndColumnNumber, other.fEndCharacterOffset); }
public void setValues(int beginLine, int beginColumn, int beginOffset,
int endLine, int endColumn, int endOffset) { fBeginLineNumber = beginLine;
fBeginColumnNumber = beginColumn;
fBeginCharacterOffset = beginOffset;
fEndLineNumber = endLine;
fEndColumnNumber = endColumn;
fEndCharacterOffset = endOffset; } // setValues(int,int,int,int)
public int getBeginLineNumber() { return fBeginLineNumber; } // getBeginLineNumber():int
public int getBeginColumnNumber() { return fBeginColumnNumber; } // getBeginColumnNumber():int
public int getBeginCharacterOffset() { return fBeginCharacterOffset; } // getBeginCharacterOffset():int
public int getEndLineNumber() { return fEndLineNumber; } // getEndLineNumber():int
public int getEndColumnNumber() { return fEndColumnNumber; } // getEndColumnNumber():int
public int getEndCharacterOffset() { return fEndCharacterOffset; } // getEndCharacterOffset():int
public boolean isSynthesized() { return false; } // isSynthesize():boolean
public String toString() { StringBuffer str = new StringBuffer();
str.append(fBeginLineNumber);
str.append(':');
str.append(fBeginColumnNumber);
str.append(':');
str.append(fBeginCharacterOffset);
str.append(':');
str.append(fEndLineNumber);
str.append(':');
str.append(fEndColumnNumber);
str.append(':');
str.append(fEndCharacterOffset);
return str.toString(); } // toString():String } // class LocationItem
```

**Fig. 4** Illustration of Data class code smell

It has an object method (Object Methods:toString(): Returns a string representation of the LocationItem object containing its location information.).

The class encapsulates data related to location information (fBeginLineNumber, BeginColumnNumber, fBeginCharacterOffset, fEndLineNumber, fEndColumnNumber, fEndCharacterOffset) for HTML events. These data members are declared as protected, indicating that they can be accessed by subclasses. The class maintains its state immutably after construction. Once the values are set using the setValues() method, they cannot be modified directly. This ensures that the location information remains consistent and prevents unintended modifications.

The class primarily focuses on storing and retrieving data, without containing any business logic or methods that manipulate the data in complex ways. It provides simple getter methods to access the data but does not include any operations that perform computations or transformations on the data. The class does not rely on external dependencies or external resources. It solely focuses on representing data related to location information, making it self-contained and independent. Thus the LocationItem class exemplifies the characteristics of a data class since according to detection rules in Table 3 and Martin Fowler [1], a class that does not perform a specific behavior on its data is called a data class.

*God Class*:

The provided code in Figures 5a & 5b defines a Java class named SimpleStorage (from SimpleStorage.java which is from webmail-0.7.10 project) which extends the FileStorage class. Since the code shown in Figure 6 is quite lengthy, the full version has been included in Appendix B, with ellipses in the figure indicating sections of the code that have been omitted.

The class imports necessary classes from Java's standard library for handling logging, file operations, XML parsing, and caching. The SimpleStorage class is declared as public and extends the FileStorage class. It contains various data members, including a static Log object for logging, constants, hash tables for resources and virtual domains (resources and vdoms), an ExpireableCache object for user caching (user_cache), and an integer for specifying the size of the user cache (user_cache_size). The class has a constructor that takes a WebMailServer object (parent) as a parameter and throws an UnavailableException. Inside the constructor, the super(parent) statement calls the constructor of the superclass (FileStorage), and the saveXMLSysData() method is invoked. The class contains several protected methods for initializing configurations (initConfig()), loading XML system data (loadXMLSysData()), and saving XML system data (saveXMLSysData()). These methods are responsible for parsing XML configuration files, handling exceptions, and logging informational messages.

```java
public class SimpleStorage extends FileStorage {
    private static Log log = LogFactory.getLog(FileStorage.class);
    public static final String user_domain_separator="|";
    protected Hashtable resources;
    protected Hashtable vdoms;
    protected ExpireableCache user_cache;
    protected int user_cache_size=100;
        public SimpleStorage(WebMailServer parent)
            throws UnavailableException {    super(parent);
        saveXMLSysData();     }
    protected void initConfig() throws UnavailableException {
        log.info("Configuration ... ");
        loadXMLSysData();
        log.info("successfully parsed XML configuration file.");     }
    protected void loadXMLSysData()
        throws UnavailableException {
        String datapath=parent.getProperty("webmail.data.path");
        String file="file://"+datapath+System.getProperty("file.separator")+"webmail.xml";
        Document root;
        .....     }
    protected void saveXMLSysData() {
        try { Document d=sysdata.getRoot();
            OutputStream cfg_out=new FileOutputStream(parent.getProperty("webmail.data.path")+
                            System.getProperty("file.separator")+ "webmail.xml");
            …
        } catch(Exception ex) {
            log.error("SimpleStorage: Error while trying to save WebMail configuration",
                 ex);       }
// (more code on the next page)
```

Fig. 5a Illustration of God class code smell

```
//This is the continuation on the second page
}
  protected void initCache() {
    super.initCache();
    cs.configRegisterIntegerKey(this,"CACHE_SIZE_USER","100","Size of the user cache");
    ...
  }
  public Enumeration getUsers(String domain) {
    String path=parent.getProperty("webmail.data.path")+System.getProperty("file.separator")+
    domain+System.getProperty("file.separator");
    File f=new File(path);
    ...
  }
  public XMLUserData createUserData(String user, String domain, String password) throws CreateUserDataException
{
    XMLUserData data;
    String template=parent.getProperty("webmail.xml.path")+
      System.getProperty("file.separator")+"userdata.xml";
    File f=new File(template);
    ...
  }
  public XMLUserData getUserData(String user, String domain, String password, boolean authenticate)
    throws UserDataException, InvalidPasswordException
  { if(authenticate) {
      auth.authenticatePreUserData(user,domain,password);
    }
    if(user.equals("")) {
      return null;
    }
    .....
    return data;
  }
  public void saveUserData(String user, String domain) {
    try {
      ...
        } else {
          log.warn("SimpleStorage: Could not write userdata ("+f.getAbsolutePath()+") for user "+user);
        }
      } else {
        log.error("SimpleStorage: Could not create path "+path+
          ". Aborting with user "+user);
      }
    } catch(Exception ex) {
      log.error("SimpleStorage: Unexpected error while trying to save user configuration "+
        "for user "+user+"("+ex.getMessage()+").", ex);
    }  }
  public void deleteUserData(String user, String domain) {
    String path=parent.getProperty("webmail.data.path")+System.getProperty("file.separator")+
      domain+System.getProperty("file.separator")+user+".xml";
    File f=new File(path);
    ...
  }
  public String toString() {
    String s="SimpleStorage:\n"+super.toString();
    s+=" - user cache: Capacity "+user_cache.getCapacity()+", Usage "+user_cache.getUsage();
    s+=", "+user_cache.getHits()+" hits, "+user_cache.getMisses()+" misses\n";
    return s;
  }
}
```

**Fig. 5b** Illustration of God class code smell

The class provides methods for managing user data, such as retrieving users for a given domain (getUsers()), creating user data objects (createUserData()), retrieving user data (getUserData()), and saving user data (saveUserData()). These methods handle operations like file I/O, XML parsing, user authentication, and caching. The deleteUserData() method deletes user data for a specified user and domain. The class logs informational messages using the Log object (log) created via LogFactory. Log messages are generated during configuration initialization, file operations, XML parsing, and user data handling. The toString() method overrides the superclass's toString() method to provide a string representation of the class, including details about the user cache's capacity, usage, hits, and misses.

The SimpleStorage class exhibits characteristics of the God Class code smell due to several reasons:

The class is quite large and contains a significant amount of code, encompassing a wide range of functionalities related to storage, caching, XML parsing, and user data management. This large size makes it difficult to understand and maintain, increasing the risk of introducing errors during development. The class has multiple responsibilities, violating the Single Responsibility Principle (SRP). It handles configuration initialization, XML data loading and saving, user data manipulation, caching, file operations, and logging. This violates the principle of encapsulating a single responsibility within a class, making the class less cohesive and harder to manage. The class is tightly coupled with various external components, such as its parent class (FileStorage), logging functionality (Log and LogFactory), and cache management (ExpireableCache). High coupling increases the class's dependencies on external components, making it less flexible and more challenging to modify or extend. The class exhibits low cohesion as it combines functionalities that may not be closely related. For example, it handles both file storage operations and XML data parsing, which are distinct concerns. This lack of cohesion makes it harder to understand and maintain the class, as its responsibilities are not well-defined or organized. According to detection rules for god class in Table 3 and Martin Fowler[1], when a class performs too many responsibilities then it is considered a god class.

Thus, from the above, it is concluded that God class and data class can not co-occur due to different symptoms depicted by them. For example, the God class performs multiple tasks, but the data class does not perform a task. Additionally, it is noted that both God class and data class detection rules are mutually exclusive, meaning that neither of them can typically occur. God class, for instance, has many complex methods, whereas data classes shouldn't, and so on.

```
private String toString(Attribute_Code code, HashSet referedLines) {
    StringBuffer buf = new StringBuffer();
    Attribute_Code.Opcode op;
    Attribute_Code.Opcode[] ops = code.codes;
    byte[][] operands;
    int ti, def, low, high, jump_count, npairs;
    String soffset;
    Attribute_LocalVariableTable.LocalVariable[] lvts = null;
    for (int i = 0; i < code.attributes_count; i++) {
        if (code.attributes[i] instanceof Attribute_LocalVariableTable) {
            lvts = ((Attribute_LocalVariableTable) code.attributes[i]).local_variable_table;
            break;
        }
    }
    if (code.code_length != 0) {
        for (int t = 0; t < ops.length; t++) {
            ...
            } else {
                buf.append(Util.padChar(config.labelPrefix + soffset, config.labelLength, ' ') + " : ");
            }
        } else {
            buf.append(config.instructionPadding);
        }
        // opcode name
        buf.append(Constants.OPCODE_NAMES[0xFF & op.opcode] + " ");
        switch (op.opcode) {
            case Constants.TABLESWITCH:
            ...
            case Constants.LOOKUPSWITCH:
            ...
            case Constants.GOTO:
            ...
            case Constants.IINC:
            ...
    for (int i = 0; i < code.attributes_count; i++) {
        if (code.attributes[i] instanceof Attribute_LocalVariableTable && ((Attribute_LocalVariableTable) code.attributes[i]).local_variable_table_length != 0) { buf.append(Constants.LINE_SEPARATER);
            buf.append(toString((Attribute_LocalVariableTable) code.attributes[i], ops));
            break; }
    // (more code on the next page)
```

**Fig. 6a** Illustration of Long method and Feature envy code smells

```
//This is the continuation on the second page
            }
            if (code.exception_table_length != 0) {
                    buf.append(Constants.LINE_SEPARATER);
                    buf.append(Constants.LINE_SEPARATER);
                    buf.append("[" + Constants.ATTRIBUTE_NAME_EXCEPTION_TABLE + ":");
                    for (int i = 0; i < code.exception_table_length; i++) {
                            buf.append(Constants.LINE_SEPARATER);
                            buf.append("start=" + config.labelPrefix + code.exception_table[i].start_pc);
                            buf.append(" , ");
                            buf.append("end=" + config.labelPrefix + code.exception_table[i].end_pc);
                            buf.append(" , ");
                            buf.append("handler=" + config.labelPrefix + code.exception_table[i].handler_pc);
                            buf.append(" , ");
                            if (code.exception_table[i].catch_type != 0) {
                                    buf.append("catch_type=" +
toString(cpl.getConstant(code.exception_table[i].catch_type)));
                            } else {
                                    buf.append("catch_type=0");
                            }
                    }
                    buf.append("]");
            }
            if (config.showLineNumber == true) {
                    for (int i = 0; i < code.attributes_count; i++) {
                            if (code.attributes[i] instanceof Attribute_LineNumberTable
                                    && ((Attribute_LineNumberTable)
code.attributes[i]).line_number_table_length != 0) {
                                    buf.append(Constants.LINE_SEPARATER);
                                    buf.append(Constants.LINE_SEPARATER);
                                    buf.append(toString((Attribute_LineNumberTable)
code.attributes[i]));
                                    break;
                            }
                    }
            }
            …
buf.append("[" + Constants.ATTRIBUTE_NAME_MAX_STACK + " : " + code.max_stack + "]");
            buf.append(Constants.LINE_SEPARATER);
buf.append("[" + Constants.ATTRIBUTE_NAME_MAX_LOCAL + " : " + code.max_locals + "]");
            return buf.toString();           }
```

**Fig. 6b** Illustration of Long method and Feature envy code smells

*Long method and Feature envy*:

Given the length of the code in Figures 6a & 6b, the complete version is provided in Appendix B, with ellipses in Figures 6a &6b indicating where portions of the code have been omitted. The toString method in Figures 6a & 6b belongs to the class SourceCodeBuilder which is from SourceCodeBuilder.java. The file SourceCodeBuilder.java belongs to jasml_0.10 project. The toString method takes an Attribute_Code object and a HashSet of referenced lines as parameters and is a method within the SourceCodeBuilder class. The SourceCodeBuilder class is a part of a Java decompiler, designed to convert Java bytecode (.class files) into equivalent Java source code. Its structure is as follows. The class is in the package com.jasml.decompiler. It imports various classes and constants from com.jasml.classes, com.jasml.helper, and other packages. The class SourceCodeBuilder has fields like cpl and config. (cpl: An instance variable of type ConstantPool, used to store the constant pool of the Java class being decompiled, and config: An instance variable of type SourceCodeBuilderConfiguration, likely used for configuration settings.)

The toString method primarily interacts with the Attribute_Code object by accessing its fields or invoking its methods extensively, it indicates a feature envy method. The method toString accesses and manipulates the internal state or behavior of the SourceCodeBuilder class less than it interacts with the Attribute_Code object or the HashSet, it necessarily exhibits feature envy code smell since if a method within a class primarily interacts with features of another class rather than its own, it may be considered to have a case of Feature Envy [1] and according to detection rules in Table 3. Developers in Object-Oriented Programming should closely associate the functionality and behavior with the data it utilizes. It seems that the presence of this code smell suggests that the method is not in the appropriate location and is more closely connected to the other class rather than the one it is currently in.

Also, the method toString is too long and contains many responsibilities hence it is depicting a long method code smell. Responsibility in software design refers to the duties or functions that a particular module, class, or function is supposed to perform. Each component in a software system should have a clearly defined responsibility. Thus, according to the detection rules specified in Table 3, the method toString is an example of long method code smell. Also, according to Martin Fowler[1], when a method contains too many lines of code it complicates the understanding of the code making it difficult to change. Such a method represents a long method code smell. Thus, the code in Figures 7a and 7b shows that long method and feature envy code smells can co-occur. Hence, the co-occurrence of Long Method and Feature Envy is a natural outcome of poor modularization and separation of concerns in a code. Modularization is the process of dividing a software system into separate, self-contained modules, each with a specific responsibility or functionality. Separation of Concerns is a design principle that states that a software system should be divided into distinct sections, each addressing a separate concern or aspect of the system's functionality. This separation improves modularity, making the system easier to develop, maintain, and scale.

Long methods tend to encapsulate multiple behaviors and responsibilities, some of which might logically belong to other classes. This leads to frequent interactions with those classes, resulting in feature envy. However, it is possible for Long Method and Feature Envy to not co-occur. A Long Method can exist without exhibiting Feature Envy if it does not interact heavily with the data or methods of other classes. Instead, it might be long due to internal complexity or poor organization within a single class. Feature Envy can occur in methods that are not long but still frequently access data or methods from other classes. These methods might be short but are misplaced, indicating that they belong to another class.

## Threats to validity

The study addresses potential risks to the legitimacy of our investigation, as outlined by Runeson [76]. The study did not consider construct validity and reliability, as they are not applicable.

*External Validity* Finally, three potential threats have been identified. (a) The analysis was limited to datasets consisting of open-source Java projects. Thus, the results can't be generalized to other programming languages. (b) The study exclusively looked at open-source projects, therefore the applicability of the findings to industrial software must be confirmed using the recommended method. (c) The study focused solely on traditional ML techniques; the results may be different for alternative ML techniques such as deep learning.

*Internal Validity* The components that have an impact on the outcomes are usually internal validity threats. The study has conducted experiments on four publicly available datasets however, experimenting with a larger number of diverse datasets will almost certainly increase the viability and conclusion of the research paper.

## 7. Conclusion

In this paper, a novel taxonomy for code smell interactions has been presented. It also identifies four possible inter-smell relationships and evaluates the effectiveness of seven popular ML models under two characteristics of different datasets: balanced and unbalanced datasets. By applying seven models to four publicly available datasets, the study identified four distinct inter-smell interactions within the same domain and its subcategories, according to the proposed classification scheme. The experimental results revealed that the Long Method (LM) consistently detects the presence of Feature Envy (FE) in both balanced and unbalanced data, whereas the God Class (GC) cannot reliably detect the presence of Data Class (DC) or vice versa. The co-occurrence of LM and FE, along with the lack of co-occurrence between GC and DC, aligns with the findings reported in the literature. To verify these results, various codes from the experimental projects were analyzed. The study suggests that recognizing inter-smell relationships using ML techniques is feasible. We believe that the proposed taxonomy offers a promising strategy in three ways: (a) It can facilitate the detection of code smells in scenarios where tools or specific knowledge for identifying them are unavailable, by leveraging insights from previously discovered code smells. (b) The taxonomy may assist in establishing benchmark datasets for code smells, addressing the current gap in the availability of standardized datasets. (c) It can aid in uncovering relationships between different code smells, enhancing our understanding of their interconnections. However, to confirm these findings, a more extensive analysis involving additional datasets and code smells is necessary. Since code smells may not share similar data distributions, future work would explore the feasibility of transfer learning for Inter $SD_iD$ (inter-smell detection within the same domain) and other discussed scenarios. The proposed taxonomy in this study focuses on interactions between two code smells; however, it can be extended to address interactions involving multiple code smells in the future.

## Appendix A

Table A.1 Illustration for each sub-scenario:
https://docs.google.com/document/d/1ov4_qs2hbIP2wdAdU1v2SgEzbUPWBqJx/edit?usp=sharing&ouid=102221464989892511063&rtpof=true&sd=true

Table A.2 Values of hyperparameters used for tuning:
https://docs.google.com/document/d/1nAJj0ifoWsHsbK3Ku5t5hXE5fsN7gFrD/edit?usp=sharing&ouid=102221464989892511063&rtpof=true&sd=true

## Appendix B

Long method& Feature Envy:
https://docs.google.com/document/d/1oNZlXnVWLxe9HuVj3-fNtiHJVhZFJjLF/edit?usp=sharing&ouid=102221464989892511063&rtpof=true&sd=true

God Class:
https://docs.google.com/document/d/1RxYUw_2Il2MXVOddLKBlyBdw-Wa3PBfG/edit?usp=sharing&ouid=102221464989892511063&rtpof=true&sd=true